\documentclass[preprintnumbers,article,amsmath,amssymb,floatfix,10pt,twocolumn,superscriptaddress]{revtex4-2}
\usepackage{amsmath}
\usepackage{eurosym}
\usepackage{epstopdf}
\usepackage{capt-of}
\usepackage{graphicx}
\usepackage{dcolumn}
\usepackage[left=0.8in, right=0.8in, top=1in, bottom=1in]{geometry}
\RequirePackage{graphicx}
\RequirePackage{float}
\RequirePackage{hyperref}
\RequirePackage{amsmath}
\RequirePackage{amssymb}
\RequirePackage{mathtools}
\usepackage{color}
\usepackage{comment}
\usepackage{xcolor}
\usepackage{multirow}
\usepackage{booktabs} 
\usepackage{amsmath} 
\usepackage{ragged2e}
\usepackage{orcidlink}
\newcommand{\be}{\begin{equation}}
	\newcommand{\ee}{\end{equation}}
\newcommand{\bea}{\begin{eqnarray}}
	\newcommand{\eea}{\end{eqnarray}}
\newcommand{\eeas}{\end{eqnarray*}}
\newcommand{\beas}{\begin{eqnarray*}}
\begin{document}
	%\color{red}
	\color{black}       %% For one column
\title{\textbf{Particle Creation and Variable Generalized Chaplygin Gas in $\mathcal{F}(\mathcal{R},\Sigma,\mathcal{T})$ Gravity}} 

		\author{ N. Myrzakulov\orcidlink{0000-0003}}\email{nmyrzakulov@gmail.com}
	\affiliation{L. N. Gumilyov Eurasian National University, Astana 010008, Kazakhstan.}
	\author{ S. H. Shekh\orcidlink{0000-0002-1932-8431}}\email{  da\_salim@rediff.com}
	\affiliation{L. N. Gumilyov Eurasian National University, Astana 010008, Kazakhstan.}
	\affiliation{Department of Mathematics, S.P.M. Science and Gilani Arts, Commerce College, Ghatanji, Yavatmal, \\Maharashtra-445301, India.}
	\affiliation{Pacif Institute of Cosmology and Selfology (PICS), Sagara, Sambalpur, Odisha 768224, India.}
	
	\author{Anirudh Pradhan\orcidlink{0000-0002-1932-8431}}
	\email{pradhan.anirudh@gmail.com}
	\affiliation{Centre for Cosmology, Astrophysics and Space Science (CCASS), GLA University, Mathura-281406, U.P., India.}
	
	\author{M. Zeyauddin\orcidlink{0000-0001-8382-8994}}
	\email[]{uddin\_m@rcjy.edu.sa}
	\affiliation{Department of General Studies (Mathematics) Jubail Industrial College, Jubail 31961, Saudi Arabia.}
	%%%%%%%%%%%%%%%%%%%%%%%%%%%%%%%%%%%%%%%
	\begin{abstract}
		
		\textbf{Abstract:} 
		
		In this work, we investigate the cosmological dynamics of a spatially flat Friedmann--Lemaître--Robertson--Walker Universe in the framework of generalized \( \mathcal{F}(\mathcal{R},\Sigma,\mathcal{T}) \) gravity by incorporating gravitationally induced particle creation together with the variable generalized Chaplygin gas scenario. The modified gravitational action depends explicitly on the Ricci scalar \( \mathcal{R} \), the matter-coupling scalar \( \Sigma \), and the trace of the energy--momentum tensor \( \mathcal{T} \), which collectively generate significant corrections to the standard cosmological evolution. The particle creation mechanism is introduced through an open thermodynamic description of the Universe. In addition, the dark sector is modeled using the variable generalized Chaplygin gas formalism. To examine the observational consistency of the model, the free parameters are constrained using the Pantheon\(^+\) Type Ia Supernova compilation together with the combined observational Hubble and Pantheon\(^+\) datasets through a statistical \(\chi^2\)-analysis. The cosmological behavior of the model is further explored through the evolution of the cosmological parameters. Furthermore, the thermodynamic properties of the model are investigated using the apparent horizon formalism. The obtained results demonstrate that the entropy evolution remains physically consistent throughout the cosmic evolution. Hence, the present \( \mathcal{F}(\mathcal{R},\Sigma,\mathcal{T}) \) gravity framework with particle creation provides a viable geometrical description of the late-time accelerated Universe and remains compatible with recent cosmological observations.\\
		
		
	\end{abstract}
	
	%\pacs{04.50+h}
	\maketitle
	
	\ %EndAName

\section{Introduction}

The discovery of the late-time accelerated expansion of the Universe through Type Ia Supernovae observations, together with Cosmic Microwave Background radiation, Baryon Acoustic Oscillations, and recent Pantheon\(^+\) data, has become one of the most important problems in modern cosmology \cite{1,2,3,4}.  Although dark energy is believed to play a dominant role in determining the late-time expansion history of the universe, its true nature and physical origin are still not completely understood. To explain the accelerated cosmic expansion, several theoretical descriptions of dark energy have been proposed, many of which involve dynamical evolution. One important candidate is the quintessence scenario, formulated through a canonical scalar field with the equation of state parameter lying in the interval $-1 < \omega < -\frac{1}{3}$. Another possibility is phantom dark energy, characterized by $\omega < -1$, which leads to the violation of the Weak Energy Condition (WEC). In addition, the quintom framework has attracted considerable attention because it allows the equation of state parameter to cross the cosmological constant boundary $\omega=-1$ during cosmic evolution \cite{5,6,7,8,9,10}.

Observational studies have also provided strong constraints on the present value of the dark energy equation of state parameter. By combining the nine-year Wilkinson Microwave Anisotropy Probe (WMAP9) observations with measurements from the Hubble parameter, Type Ia Supernovae, Cosmic Microwave Background radiation, and Baryon Acoustic Oscillations, the present-day value was estimated as $\omega_0 \approx -1.084 \pm 0.063$ \cite{11}. Later, the \textit{Planck} mission improved these constraints and reported $\omega_0 = -1.006 \pm 0.0451$ in 2015 , while subsequent analyses from the 2018 \textit{Planck} data yielded a more accurate estimate of $\omega_0 = -1.028 \pm 0.032$ \cite{12}.

Although the standard \(\Lambda CDM\) cosmological model successfully explains the observed expansion history of the Universe, it still suffers from several theoretical difficulties such as the cosmological constant problem, fine-tuning problem, and coincidence problem \cite{13,14}. These limitations have motivated the construction of several modified theories of gravity as possible alternatives to Einstein’s General Relativity. In recent years, theories such as \(f(R)\), \(f(T)\), \(f(Q)\), and \(f(R,T)\) gravity \cite{15,16,17,18,19,20,21,22,23,24,25,26} have received considerable attention because they provide a geometrical explanation for dark energy and late-time cosmic acceleration without necessarily introducing exotic matter fields. Among these generalized frameworks, curvature--matter coupling theories play an important role in describing the interaction between matter and geometry at cosmological scales.

As, the well-known $f(R,T)$ theory has been generalized by incorporating the additional quantity $\Sigma = T_{\mu\nu}T^{\mu\nu}$, leading to the development of the $\mathcal{F}(\mathcal{R}, \Sigma ,\mathcal{T})$ gravity model \cite{27,28}. The presence of the term $\Sigma$ introduces nonlinear interactions between matter and geometry, which can play an important role in describing various astrophysical and cosmological effects. This extended framework has the potential to explain the accelerated phases of the universe, both at early and late times, as well as phenomena such as galactic rotation curves, without necessarily invoking dark matter or dark energy components. Motivated by these developments, the generalized $\mathcal{F}(\mathcal{R}, \Sigma ,\mathcal{T})$ gravity theory has emerged as an interesting extension of modified gravity models. In this framework, the gravitational action depends not only on the Ricci scalar \(R\) and the trace of the energy--momentum tensor \(T\), but also on an additional matter-coupling scalar \(\Sigma\), which represents nonlinear matter corrections. The presence of these additional coupling terms modifies the effective gravitational interaction and allows richer cosmological dynamics compared with standard Einstein gravity.

Another important mechanism capable of explaining the accelerated expansion of the Universe is gravitationally induced particle creation \cite{29,30}. In particle creation cosmology, the Universe is treated as an open thermodynamic system where continuous matter production generates an effective negative pressure. Such a mechanism can naturally drive the accelerated expansion of the Universe without requiring a pure cosmological constant dominated scenario.

To describe the dark sector evolution of the Universe, the variable generalized Chaplygin gas model has also attracted significant attention in recent years. The VGCG model smoothly interpolates between the matter-dominated phase and the late-time accelerated epoch, thereby providing a unified description of dark matter and dark energy. Consequently, the combined study of particle creation, generalized Chaplygin gas, and modified gravity may provide a more complete description of cosmic evolution \cite{31,32}.

From the observational point of view, it is necessary to test the physical viability of cosmological models using recent astrophysical datasets. In this regard, observational Hubble data and the Pantheon\(^+\) Type Ia Supernova compilation provide strong constraints on the expansion history of the Universe and help determine the allowed parameter space of the theoretical model.

Furthermore, several cosmological diagnostic tools such as the deceleration parameter, statefinder parameters, and equation of state parameter are useful for distinguishing different dark energy models and understanding the dynamical behavior of the Universe. In addition, thermodynamic analysis plays an important role in examining the physical consistency and stability of modified gravity theories during cosmic evolution.

In the present work, we investigate particle creation cosmology in the framework of $\mathcal{F}(\mathcal{R}, \Sigma ,\mathcal{T})$ gravity by considering a variable generalized Chaplygin gas scenario in a spatially flat FLRW Universe. The corresponding cosmological model is constrained using Pantheon\(^+\) and combined observational Hubble datasets. We further analyze the physical behavior of the model through the evolution of the deceleration parameter, statefinder diagnostics, effective equation of state parameter, and thermodynamic properties of the Universe.
\section{$\mathcal{F}(\mathcal{R}, \Sigma ,\mathcal{T})$ Gravity Framework}

In recent years, extensions involving nonlinear matter corrections have emerged as promising alternatives for describing both early and late cosmic evolution.

Motivated by these developments, we consider a generalized gravitational framework in which the gravitational Lagrangian depends not only on the Ricci scalar curvature but also on additional matter-coupling contributions. In this approach, the gravitational action is expressed as \cite{33}
\begin{align}
	\mathcal{S}
	=
	\frac{1}{16\pi}
	\int
	\sqrt{-g}
	\left[
	\mathcal{F}(\mathcal{R}, \Sigma ,\mathcal{T})
	+
	\mathcal{L}_{m}
	\right]
	d^{4}x,
	\label{1}
\end{align}
where \(g\) denotes the determinant of the metric tensor \(g_{\mu\nu}\), \(\mathcal{L}_{m}\) represents the matter Lagrangian density, and \(\mathcal{F}(\mathcal{R}, \Sigma ,\mathcal{T})\) is an arbitrary analytical function of the Ricci scalar \(\mathcal{R}\), the trace of the energy--momentum tensor \(\mathcal{T}\), and an additional matter-coupling scalar \( \Sigma \).

The quantity \( \Sigma \) is introduced in order to encode nonlinear interactions between matter and geometry. Following the generalized matter-coupling formalism, we define
\begin{align}
	\Sigma _{\mu\nu}
	=
	\chi\,\Theta_{\mu\nu},
	\label{2}
\end{align}
where \(\chi\) is a dimensionless coupling parameter controlling the strength of the interaction between attractive and repulsive gravitational sectors, while \(\Theta_{\mu\nu}\) characterizes higher-order matter corrections. Consequently, the effective geometrical tensor may be written as
\begin{align}
	\mathcal{B}_{\mu\nu}
	=
	\mathcal{R}_{\mu\nu}
	+
	\Sigma _{\mu\nu}.
	\label{3}
\end{align}

The action in Eq. \eqref{1} generalizes several well-known modified gravity theories. In the limit \( \Sigma  \rightarrow 0\), the theory reduces to the standard \(f(R,T)\) gravity framework whereas when both \( \Sigma \) and \(\mathcal{T}\) vanish, Einstein gravity is naturally recovered.

Variation of the action \eqref{1} with respect to the metric tensor \(g_{\mu\nu}\) yields the modified gravitational field equations
\begin{align}
	&
	\mathcal{F}_{\mathcal{R}}
	\mathcal{R}_{\mu\nu}
	-
	\frac{1}{2}
	g_{\mu\nu}
	\mathcal{F}
	+
	\left(
	g_{\mu\nu}\Box
	-
	\nabla_{\mu}\nabla_{\nu}
	\right)
	\mathcal{F}_{\mathcal{R}}
	-
	\Sigma _{\mu\nu}
	\mathcal{F}_{ \Sigma }
	\nonumber\\
	&
	=
	8\pi
	T_{\mu\nu}
	-
	\mathcal{F}_{\mathcal{T}}
	\left(
	T_{\mu\nu}
	+
	\Xi_{\mu\nu}
	\right),
	\label{4}
\end{align}
where $\mathcal{F}_{\mathcal{R}}
=
\frac{\partial \mathcal{F}}{\partial \mathcal{R}},
\qquad
\mathcal{F}_{ \Sigma }
=
\frac{\partial \mathcal{F}}{\partial  \Sigma },
\qquad
\mathcal{F}_{\mathcal{T}}
=
\frac{\partial \mathcal{F}}{\partial \mathcal{T}}$, and \(\Box=\nabla^{\mu}\nabla_{\mu}\) denotes the d'Alembert operator. The matter energy--momentum tensor is defined through the usual variational prescription
\begin{align}
	T_{\mu\nu}
	=
	(\rho+p)u_{\mu}u_{\nu}
	+
	pg_{\mu\nu},
	\label{5}
\end{align}
where \(\rho\) and \(p\) correspond respectively to the energy density and isotropic pressure of the cosmic fluid, while \(u_{\mu}\) is the four-velocity satisfying \(u_{\mu}u^{\mu}=-1\).

For the present cosmological analysis, we adopt the linear functional form
\begin{align}
	\mathcal{F}(\mathcal{R}, \Sigma ,\mathcal{T})
	=
	\mathcal{R}
	+
	\Sigma 
	+
	2\lambda \mathcal{T},
	\label{6}
\end{align}
where \(\lambda\) is a constant matter-coupling parameter. This choice represents the simplest nontrivial extension of Einstein gravity containing both trace and nonlinear matter corrections.

Substituting Eq. \eqref{6} into the general field equations \eqref{4}, we obtain
\begin{align}
	\mathcal{R}_{\mu\nu}
	+
	\Sigma _{\mu\nu}
	-
	\frac{1}{2}
	g_{\mu\nu}
	(\mathcal{R}+ \Sigma )
	=
	(8\pi+2\lambda)
	T_{\mu\nu}
	+
	\lambda g_{\mu\nu}
	(\mathcal{T}+2p).
	\label{7}
\end{align}

Equation \eqref{7} governs the cosmological dynamics in the \(\mathcal{F}(\mathcal{R}, \Sigma ,\mathcal{T})\) gravity framework. The additional coupling terms generated through \( \Sigma \) and \(\mathcal{T}\) modify the effective gravitational interaction and provide richer cosmological dynamics compared with standard general relativity.

The inclusion of nonlinear matter corrections plays an important role in the present formalism. Unlike conventional \(f(R)\) models, where the geometrical sector alone drives cosmic acceleration, the present theory incorporates direct couplings between matter and geometry. Such couplings naturally emerge in high-energy regimes and may significantly affect the cosmic dynamics during different evolutionary stages of the Universe. Moreover, the additional degrees of freedom introduced through \( \Sigma \) allow the model to explain the transition from a decelerated matter-dominated phase to the present accelerated epoch without requiring exotic scalar fields or phantom fluids.

%%%%%%%%%%%%%%%%%%%%%%%%%%%%%%%%%%%%%%%%%%%%%%%%%%%%%%%%%%%%%%%%%%%%%%%

\section{Cosmological Geometry and Field Equations}

To investigate the cosmological consequences of the modified gravity model introduced above, we consider a homogeneous and isotropic spacetime described by the spatially flat Friedmann--Lemaître--Robertson--Walker (FLRW) metric,
\begin{align}
	ds^{2}
	=
	-dt^{2}
	+
	a^{2}(t)
	\left[
	dx^{2}
	+
	dy^{2}
	+
	dz^{2}
	\right],
	\label{8}
\end{align}
where \(a(t)\) denotes the cosmic scale factor. The Hubble expansion parameter is defined in the standard manner as $	H=\frac{\dot{a}}{a}$, where an overdot represents differentiation with respect to cosmic time \(t\). For the FLRW geometry, the Ricci scalar takes the form $\mathcal{R}=6\left(\dot{H}+2H^{2}\right)$. Substituting the metric \eqref{8} into the field equations \eqref{7}, we obtain the modified Friedmann equations governing the cosmological evolution:
\begin{align}
	3(1-\chi)^{2}H^{2}
	=
	\lambda_{1}\rho
	-
	\lambda_{2}p,
	\label{9}
\end{align}
\begin{align}
	3(\chi-1)
	\left(
	\dot{H}
	+
	H^{2}
	\right)
	=
	\lambda_{3}\rho
	+
	\lambda_{4}p,
	\label{10}
\end{align}
where the coefficients are defined as $\lambda_{1}=\pi(8+3\lambda),
\quad
\lambda_{2}=\pi \lambda,
\quad
\lambda_{3}=4\pi,
\quad
\lambda_{4}=4\pi(3+\lambda)$.
Equations \eqref{9} and \eqref{10} constitute a coupled system relating the thermodynamic quantities \(\rho\) and \(p\) with the Hubble parameter and its derivative.

%%%%%%%%%%%%%%%%%%%%%%%%%%%%%%%%%%%%%%%%%%%%%%%%%%%%%%%%%%%%%%%%%%%%%%%
An explicit expressions for the cosmic energy density and pressure from the system (\ref{9}) and (\ref{10}), written as
\begin{align}
	\rho=\frac{3\lambda_{4}(1-\chi)^{2}H^{2}+3\lambda_{2}(\chi-1)(\dot{H}+H^{2})}{\lambda_{1}\lambda_{4}+\lambda_{2}\lambda_{3}}.
	\label{11}
\end{align}
\begin{align}
	p=\frac{3\lambda_{1}(\chi-1)(\dot{H}+H^{2})-3\lambda_{3}(1-\chi)^{2}H^{2}}{\lambda_{1}\lambda_{4}	+\lambda_{2}\lambda_{3}}.
	\label{12}
\end{align}

The expressions \eqref{11} and \eqref{12} represent the generalized energy density and pressure relations in \(\mathcal{F}(\mathcal{R}, \Sigma ,\mathcal{T})\) gravity. These equations clearly demonstrate that the cosmic evolution depends not only on the Hubble parameter but also on the nonlinear matter-coupling parameters \(\chi\) and \(\lambda\), thereby leading to modified cosmological dynamics compared with standard Einstein gravity.

%%%%%%%%%%%%%%%%%%%%%%%%%%%%%%%%%%%%%%%%%%%%%%%%%%%%%%%%%%%%%%%%%%%%%%%

\subsection{Matter Creation Cosmology}

In the present framework, the Universe is treated as an open thermodynamic system in which gravitationally induced particle production may occur continuously during cosmic evolution \cite{Prigogine1989,Calvao1992,Lima2014}. Such a mechanism modifies the standard conservation law and introduces an effective negative pressure capable of driving accelerated expansion.

The particle number balance equation is written as
\begin{align}
	\nabla_{\mu}N^{\mu}
	=
	\dot{n}
	+
	3Hn
	=
	n\Gamma,
	\label{13}
\end{align}
where \(n\) denotes the particle number density and \(\Gamma\) is the particle creation rate.

For adiabatic matter creation, the creation pressure takes the form
\begin{align}
	p_{c}
	=
	-
	\frac{\rho+p}{3H}\Gamma.
	\label{14}
\end{align}

Assuming the phenomenological relation
\begin{align}
	\Gamma
	=
	3\zeta H,
	\label{15}
\end{align}
where \(\zeta\) is the matter creation parameter, the conservation equation becomes
\begin{align}
	\dot{\rho}_{m}
	+
	3H(1-\zeta)\rho_{m}
	=
	0.
	\label{16}
\end{align}

Integrating Eq. \eqref{16}, we obtain
\begin{align}
	\rho_{m}=\rho_{m0}(1+z)^{3(1-\zeta)},	\label{17}
\end{align}
where \(\rho_{m0}\) is the present value of the matter density.

%%%%%%%%%%%%%%%%%%%%%%%%%%%%%%%%%%%%%%%%%%%%%%%%%%%%%%%%%%%%%%%%%%%%%%%

\subsection{Variable Generalized Chaplygin Gas}

To describe the dark sector of the Universe, we employ the variable generalized Chaplygin gas (VGCG) model characterized by the equation of state \cite{Guo2007,Yang2007,Lu2009}
\begin{align}
	p_{vg}
	=
	-
	\frac{\mathcal{A}(a)}
	{\rho_{vg}^{\alpha}},
	\label{18}
\end{align}
where $\mathcal{A}(a)=\mathcal{A}_{0}a^{-m}$. Substituting Eq. \eqref{18} into the conservation equation  $\dot{\rho}_{vg}+3H(\rho_{vg}+p_{vg})=0$, we obtain
\begin{align}
	\rho_{vg}=\rho_{vg0}\left[A_{s}(1+z)^{m}+(1-A_{s})(1+z)^{3(1+\alpha)}\right]^{\frac{1}{1+\alpha}},
	\label{19}
\end{align}
where $A_{s}=\frac{3(1+\alpha)}{3(1+\alpha)-m}\frac{\mathcal{A}_{0}}{\rho_{vg0}^{1+\alpha}}$. The VGCG model smoothly interpolates between a matter-dominated epoch at high redshift and a dark-energy-dominated accelerated phase at late times, making it a suitable candidate for unified dark sector cosmology. Considering the total energy density as $\rho=\rho_{vg}+\rho_{m}$, we now derive the corresponding expression for the Hubble parameter \(H(z)\). From the equations (\ref{17}) and (\ref{19}), the total energy density becomes
\begin{align}
	\rho=
	\rho_{m0}(1+z)^{3(1-\zeta)}
	+\nonumber\\	\rho_{vg0}
	\left[
	A_{s}(1+z)^{m}
	+(1-A_{s})(1+z)^{3(1+\alpha)}
	\right]^{\frac{1}{1+\alpha}}.
	\label{20}
\end{align}

Substituting Eq.~\eqref{20} into Eq.~\eqref{11} and using the relation between cosmic time and redshift. Thus, the differential equation governing the Hubble parameter becomes
\begin{small}
	\begin{align}
		(1+z)H\frac{dH}{dz}-\frac{\lambda_{4}(1-\chi)^{2}+\lambda_{2}(\chi-1)}{\lambda_{2}(\chi-1)}H^{2}=\nonumber\\ \left(-\frac{\lambda_{1}\lambda_{4}+\lambda_{2}\lambda_{3}}{3\lambda_{2}(\chi-1)}\right)\times \Bigg[\rho_{m0}(1+z)^{3(1-\zeta)}+ \nonumber\\ \rho_{vg0}\left(
		A_{s}(1+z)^{m}+(1-A_{s})(1+z)^{3(1+\alpha)}\right)^{\frac{1}{1+\alpha}}\Bigg].\label{21}
	\end{align}
\end{small}
Defining \(Y(z)=H^{2}(z)\), Eq.~\eqref{21} can be rewritten in the standard linear differential form as
\begin{align}\label{22}
	\frac{dY}{dz}+P(z)Y=Q(z),
\end{align}
where
$P(z)=
-\frac{
	2\left[
	\lambda_{4}(1-\chi)^{2}
	+\lambda_{2}(\chi-1)
	\right]
}{
	\lambda_{2}(\chi-1)(1+z)
}$,
and
$\\
Q(z)=
-\frac{
	2(\lambda_{1}\lambda_{4}+\lambda_{2}\lambda_{3})
}{
	3\lambda_{2}(\chi-1)(1+z)
}
\Bigg[
\rho_{m0}(1+z)^{3(1-\zeta)}
\nonumber\\
+
\rho_{vg0}
\left(
A_{s}(1+z)^{m}
+
(1-A_{s})(1+z)^{3(1+\alpha)}
\right)^{\frac{1}{1+\alpha}}
\Bigg].
$

Introducing the parameter $\eta=\frac{2\left[\lambda_{4}(1-\chi)^{2}+\lambda_{2}(\chi-1)\right]}{\lambda_{2}(\chi-1)}$,
the equation (\ref{21}) reduces to a linear differential equation of the form 
\begin{align}
	\frac{dY}{dz}
	-
	\frac{\eta}{1+z}Y
	=
	Q(z).\label{23}
\end{align}
whose integrating factor is $\mu(z)=(1+z)^{-\eta}$ and the corresponding solution for the Hubble parameter  as
\begin{align}\label{24}
	H^{2}(z)
	=
	(1+z)^{\eta}
	\left[
	C+I_{1}+I_{2}
	\right],
\end{align}
where \(C\) is the integration constant, while \(I_{1}\) and \(I_{2}\) denote the matter and viscous generalized Chaplygin gas contributions, respectively. The above equation of Hubble parameter admits an exact analytical solution. After solving, consequently the Hubble parameter is expressed in the compact form as
\begin{widetext}
	\begin{align}
		H(z)=H_0\sqrt{A(1+z)^\eta-B(1+z)^{3(1-\zeta)}-C(1+z)^m-D(1+z)^3},\label{25}
	\end{align}
\end{widetext}
where $A=1+\Omega_{m0}+\Omega_{vg0}\left(\frac{A_s}{m-\eta}+\frac{1-A_s}{3-\eta}\right)$, $B=\Omega_{m0}$, $C=\Omega_{vg0}\left(\frac{A_s}{m-\eta}\right)$, and $D=\Omega_{vg0}
\left(\frac{1-A_s}{3-\eta}\right)$.
The reduced expression given by Eq.~\eqref{25} is particularly convenient for observational analysis. In the subsequent section, we constrain the model parameters
$$(H_0,A,B,C,D,\eta,\zeta,m)$$, using recent observational datasets including Cosmic Chronometers (CC), Baryon Acoustic Oscillations (BAO), and Pantheon supernova compilations.
\section{Cosmological Viability Through Observational Analysis}\label{IV}

In order to examine the physical reliability and observational consistency of the proposed cosmological framework, we confront the theoretical predictions of the model with recent astrophysical observations. The accelerated expansion history of the Universe can be effectively tested through direct measurements of the Hubble parameter and luminosity distance observations from Type Ia Supernovae. Therefore, to place meaningful constraints on the free parameters of the model, we employ two independent observational probes, namely the observational Hubble data (OHD) and the Pantheon+ compilation of Type Ia Supernovae data.

The estimation of the model parameters is performed through a standard statistical likelihood analysis based on the minimization of the chi-square ($\chi^2$) function. Such a statistical procedure enables us to determine the best-fit values of the cosmological parameters and to evaluate the compatibility of the present theoretical construction with observational evidence. The joint analysis involving both datasets provides stronger constraints and improves the robustness of the cosmological predictions.

%%%%%%%%%%%%%%%%%%%%%%%%%%%%%%%%%%%%%%%%%%%%%%%%%%%%%%%%%%%%%

\subsubsection{Constraints from Observational Hubble Data (OHD)}

The observational Hubble dataset constitutes one of the most direct probes of the cosmic expansion history. In the present analysis, we utilize a compilation of $77$ measurements of the Hubble parameter $H(z)$ distributed over the redshift interval $0.07 \leq z \leq 1.965$ \cite{a77}. These measurements are mainly obtained through the differential age technique, commonly known as the cosmic chronometer approach \cite{a78,a79}. This method determines the Hubble parameter from the relative age evolution of passively evolving galaxies without assuming any specific cosmological model. A detailed discussion regarding the construction and compilation of these data can be found in \cite{a80}.

The observational Hubble parameter is related to the expansion rate of the Universe through
\begin{equation}
	H(z)=-\frac{1}{1+z}\frac{dz}{dt},\label{28}
\end{equation}
which directly characterizes the dynamical evolution of the cosmic scale factor. Since the Hubble parameter enters explicitly into the field equations and cosmological observables, the OHD dataset serves as an efficient tool for constraining the free parameters of the model.

To determine the best-fit values of the model parameters, we employ the chi-square minimization method. The corresponding chi-square function is defined as
\begin{equation}\label{29}
	\chi^{2}\left(H_{0}, Cons. \right)=\sum_{i=1}^{77} {\frac{\left(H_{th}(z_i,Cons.)-H_{ob}(z_i)\right)^2}{\sigma^2_{i}}},
\end{equation}
where $H_{th}(z_i)$ represents the theoretical prediction of the Hubble parameter at redshift $z_i$, while $H_{ob}(z_i)$ denotes the corresponding observed value. The quantity $\sigma_i$ corresponds to the standard deviation associated with each observational point. The parameters $Cons. \approx H_0,A,B,C,D,\eta,\zeta$ and $m$ denote the free model parameters to be constrained through the statistical analysis.

The minimization of the above estimator yields the optimum parameter space for which the theoretical model exhibits the closest agreement with observational measurements. Moreover, confidence contours generated from the likelihood distribution provide additional information regarding the allowed parameter regions at different confidence levels.

%%%%%%%%%%%%%%%%%%%%%%%%%%%%%%%%%%%%%%%%%%%%%%%%%%%%%%%%%%%%%

\subsubsection{Pantheon+ Type Ia Supernova Compilation}

Type Ia Supernovae are among the most reliable cosmological distance indicators and play a crucial role in studying the late-time acceleration of the Universe. In the present work, we employ the recently released Pantheon+ compilation of Supernovae Ia data in the redshift interval $0.001 < z < 2.26$ \cite{Brout/2022a}. This dataset represents one of the most comprehensive and statistically refined collections of SN Ia observations currently available. The Pantheon+ sample combines observations from $18$ different astronomical surveys and provides a homogeneous dataset with improved calibration and reduced systematic uncertainties. The observed supernova light curves are standardized using the SALT2 light-curve fitting model \cite{Brout/2022b}, which ensures uniformity in the determination of distance moduli across the entire dataset.

The theoretical analysis of SN Ia observations is based on the distance modulus relation
\begin{equation}
	\mu(z)=m-M_b=\mu_0+5\log D_L(z),\label{30}
\end{equation}
where $m$ and $M_b$ denote the apparent and absolute magnitudes, respectively. Here, $D_L(z)$ represents the luminosity distance, while $\mu_0$ is the nuisance parameter associated with the Hubble constant. For a spatially homogeneous and isotropic Universe, the luminosity distance is expressed as
\begin{equation}
	D_L(z)=(1+z)\int_{0}^{z}\frac{dz'}{H(z')},\label{31}
\end{equation}
which directly depends upon the Hubble expansion history predicted by the cosmological model. The chi-square estimator corresponding to the Pantheon+ dataset is given by
\begin{equation}
	\label{32}
	\chi^{2}_{Pantheon+} = (m_{obs} - m_{th})^{T}C^{-1}_{Pantheon+}(m_{obs} - m_{th}),
\end{equation}
where $m_{obs}$ and $m_{th}$ denote the observed and theoretical apparent magnitudes, respectively, and $C^{-1}_{Pantheon+}$ represents the inverse covariance matrix associated with the Pantheon+ compilation. The covariance matrix incorporates both statistical and systematic uncertainties, thereby allowing a more accurate and reliable estimation of the cosmological parameters.

The Pantheon+ dataset is particularly significant because of its broad redshift coverage and high statistical precision, which make it highly effective in constraining the late-time behavior of dark energy models and modified gravity scenarios.

%%%%%%%%%%%%%%%%%%%%%%%%%%%%%%%%%%%%%%%%%%%%%%%%%%%%%%%%%%%%%

\subsubsection{Joint Statistical Analysis}

To obtain more stringent and reliable bounds on the cosmological parameters, we perform a combined statistical analysis involving both the OHD and Pantheon+ datasets. The total chi-square function is constructed as the sum of the individual chi-square estimators corresponding to each dataset:
\begin{equation}\label{33}
	\chi^{2}_{joint} = \chi^{2}_{OHD} + \chi^{2}_{Pantheon+}.
\end{equation}

The minimization of the combined chi-square function provides the global best-fit values of the model parameters. The joint analysis significantly reduces parameter degeneracies and improves the overall confidence in the observational viability of the model. Consequently, the obtained constrained parameter space allows us to investigate the cosmological dynamics of the proposed framework in accordance with recent observational evidence.
\begin{figure*}
	\centering
	\includegraphics[scale=0.450]{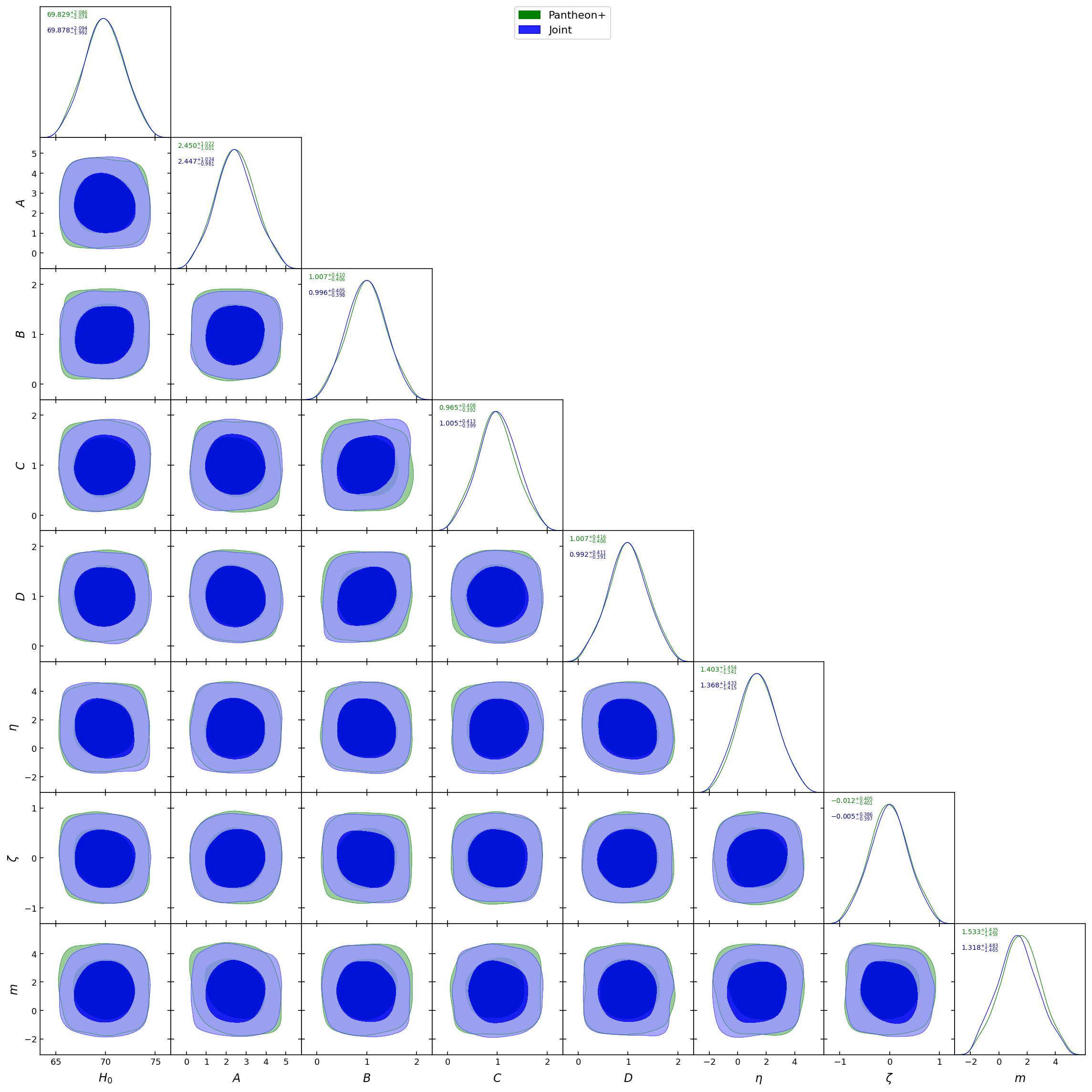}
	\caption{One-dimensional marginalized distribution, and $2D$ contours with $1\sigma$ and $2\sigma$ confidence levels for this model restricted with combined Pantheon+ and joint data sets.}\label{OC}
\end{figure*}
%%%%%%%%%%%%%%%%%%%%%%%%%%%%%%%%%%%%%%%%%%%%%%%%%%%%%%%%%%%%%%%%%%%%%%%%%	\\
Fig.~\ref{OC} display the $1D$ marginalized posterior distributions together with the corresponding $2D$ confidence contours for the free parameters of the present cosmological model. The contours represent the $1\sigma$ and $2\sigma$ confidence levels obtained from the statistical analysis of the Pantheon+ compilation of SN Ia data and the combined OHD + Pantheon+ dataset, respectively. These contour plots provide a graphical representation of the allowed parameter space and illustrate the degree of correlation among the model parameters. The comparatively narrow contours in the joint analysis indicate that the inclusion of the OHD dataset significantly improves the parameter constraints and reduces the associated uncertainties. The best-fit values obtained from the Pantheon+ the combined OHD + Pantheon+ analysis are given in the following table \ref{tab1} :
\begin{table*}
	\centering
	\renewcommand{\arraystretch}{1.8}
	\setlength{\tabcolsep}{18pt}
	\caption{Best-fit values of the model parameters obtained from the Pantheon+ and combined OHD + Pantheon+ datasets at $1\sigma$ and $2\sigma$ confidence level.}
	\label{tab1}
	\begin{tabular}{|c|c|c|}
		\hline \hline
		\textbf{Parameter} & \textbf{Pantheon+} & \textbf{OHD + Pantheon+} \\
		\hline \hline
		
		$H_0$ 
		& $69.829^{+2.086}_{-2.074}\,\text{km/s/Mpc}$ 
		& $69.878^{+2.094}_{-1.992}\,\text{km/s/Mpc}$ \\
		\hline
		
		$A$ 
		& $2.450^{+1.022}_{-1.001}$ 
		& $2.447^{+1.024}_{-0.981}$ \\
		\hline
		
		$B$ 
		& $1.007^{+0.410}_{-0.406}$ 
		& $0.996^{+0.405}_{-0.398}$ \\
		\hline
		
		$C$ 
		& $0.965^{+0.408}_{-0.392}$ 
		& $1.005^{+0.413}_{-0.399}$ \\
		\hline
		
		$D$ 
		& $1.007^{+0.416}_{-0.406}$ 
		& $0.992^{+0.411}_{-0.391}$ \\
		\hline
		
		$\eta$ 
		& $1.403^{+1.454}_{-1.041}$ 
		& $1.368^{+1.433}_{-1.415}$ \\
		\hline
		
		$\zeta$ 
		& $-0.012^{+0.405}_{-0.401}$ 
		& $-0.005^{+0.386}_{-0.397}$ \\
		\hline
		
		$m$ 
		& $1.553^{+1.435}_{-1.459}$ 
		& $1.318^{+1.483}_{-1.465}$ \\
		\hline
		
		$M$ 
		& $23.4873^{+0.6330}_{-0.6109}$ 
		& $23.5717^{+0.6087}_{-0.6035}$ \\
		\hline
		
	\end{tabular}
\end{table*}
The obtained values indicate that the present cosmological model remains observationally viable within the allowed confidence regions of the employed datasets. In particular, the Hubble constant estimated from the Pantheon+ and the joint dataset respectively as,
$$H_{0_{Pantheon+}}=69.829^{+2.086}_{-2.074}\,\text{km/s/Mpc}$$ and 
$$H_{0_{joint}}=69.878^{+2.094}_{-1.992}\,\text{km/s/Mpc}$$
lies well within the range reported by several recent cosmological observations and demonstrates good consistency with late-time observational probes. The inclusion of the OHD dataset alongside the Pantheon+ sample slightly tightens the bounds on the model parameters, thereby enhancing the statistical reliability of the analysis. It is worth emphasizing that the estimated values of the parameters $B$, $C$, and $D$ remain close to unity for both datasets, suggesting a stable contribution of the corresponding terms in the proposed Hubble parametrization throughout the cosmic evolution. Furthermore, the positive values of the parameters $\eta$ and $m$ indicate that the model successfully accommodates a dynamically evolving Universe with a smooth transition from decelerated expansion to the presently accelerating phase. The nuisance parameter $M$, associated with the absolute magnitude calibration in the Pantheon+ compilation, also attains consistent values in both analyses, supporting the internal consistency of the observational fitting procedure. Moreover, the agreement between the Pantheon+ and the combined OHD + Pantheon+ analyses demonstrates the robustness of the model against different observational probes.

Since the Pantheon+ and combined OHD + Pantheon+ datasets provide comparatively stronger and more constrained parameter spaces, the best-fit values obtained from these two analyses are subsequently employed in the investigation of the physical and kinematical behavior of the cosmological model. In particular, these constrained values are used to analyze the evolution of various cosmological quantities such as the deceleration parameter, equation of state parameter, energy density, pressure, and other diagnostic parameters describing the dynamical evolution of the Universe.
\section{Physical behavior of the model}
\subsubsection{Deceleration parameter}
The evolution of the deceleration parameter provides important information regarding the expansion history of the Universe and the transition from an early decelerated epoch to the presently observed accelerated phase. For the obtained cosmological model, the deceleration parameter is expressed as
\begin{widetext}
\begin{equation}
	q(z)=
	-\frac{
		(-z-1)
		\left[
		A\eta (1+z)^{\eta-1}
		+
		3B(1-\xi)(1+z)^{3(1-\xi)-1}
		+
		Cm(1+z)^{m-1}
		+
		3D(1+z)^2
		\right]
	}
	{
		2\left[
		A(1+z)^\eta
		+
		B(1+z)^{3(1-\xi)}
		+
		C(1+z)^m
		+
		D(1+z)^3
		\right]
	}
	-1.
	\label{34}
\end{equation}
\end{widetext}

\begin{figure}[ht]
	\centering
	\includegraphics[scale=0.6]{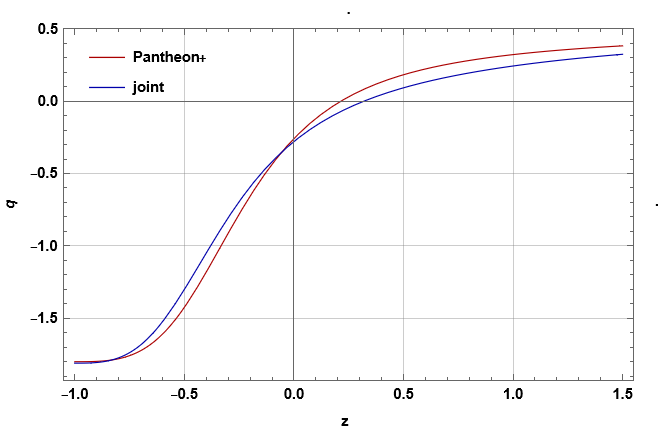}
	\caption{Evolution of the deceleration parameter \(q(z)\) as a function of redshift \(z\) for the constrained values of the model parameters using the Pantheon\(^+\) and joint \((CC+\mathrm{Pantheon}^+)\) datasets}.\label{q}
\end{figure}

The graphical evolution of \(q(z)\) for both the Pantheon\(^+\) and the joint \((CC+\text{Pantheon}^+)\) datasets is shown in Fig.~\ref{q}. The corresponding best-fit values of the model parameters are summarized in the table (\ref{tab1}). For the Pantheon\(^+\) dataset and the joint dataset, the present value of the deceleration parameter at \(z=0\) one respectively obtains $q_0 \approx -0.27$ and $q_0 \approx -0.26$. These negative values clearly indicate that the present Universe is undergoing an accelerated phase of expansion. The obtained results are consistent with several recent observational investigations based on supernovae, baryon acoustic oscillations, and cosmic chronometer measurements, which generally predict the present value of the deceleration parameter within the range $-1<q_0<0$.
In particular, observational analyses within the standard \(\Lambda CDM\) framework usually provide values close to $q_0 \simeq -0.5 \; \text{to} \; -0.6$, depending on the adopted datasets and parametrization scheme. The values obtained in the present investigation remain compatible with the accelerated expansion scenario, although the acceleration predicted by the model is comparatively milder than the standard cosmological constant dominated model. The graphical behavior further reveals that at high redshift (\(z>0\)) the deceleration parameter remains positive, corresponding to a matter-dominated decelerating Universe. As the Universe evolves toward lower redshift, the deceleration parameter gradually decreases and crosses the transition line $q(z_t)=0$, indicating the cosmic transition from deceleration to acceleration. Further, from the graphical analysis, the transition redshift is approximately obtained as $z_t \approx 0.22$ for the Pantheon\(^+\) dataset, and $z_t \approx 0.30$ for the joint dataset. These transition values are physically acceptable and remain close to several observationally reconstructed transition redshifts reported in the literature, where the transition generally occurs within the interval $0.3 \lesssim z_t \lesssim 1$. The figure also shows that in the far future region \((z\rightarrow -1)\), the deceleration parameter approaches significantly negative values. This indicates that the accelerated expansion continues in the future evolution of the Universe. Near \(z=-1\), the Pantheon\(^+\) curve approaches $q \approx -1.8$, while the joint dataset predicts $q \approx -1.7$. Such behavior suggests a strongly accelerating late-time phase driven by the effective dark energy component generated through the modified gravitational corrections and viscous Chaplygin gas contribution.

Overall, the evolution of the deceleration parameter successfully reproduces the essential cosmological sequence consisting of $\textbf{Decelerated phase}
\quad \longrightarrow \quad
\textbf{Transition epoch}
\quad \longrightarrow \quad
\textbf{Accelerated phase}$, which is one of the fundamental requirements for any viable cosmological model describing the late-time Universe.

\subsubsection{Statefinder Diagnostic Analysis}

To further investigate the dynamical behavior of the proposed cosmological model and distinguish it from other dark energy scenarios, we employ the statefinder diagnostic pair \((r,s)\). The statefinder formalism provides a powerful geometrical method for analyzing the expansion dynamics of the Universe beyond the Hubble and deceleration parameters. Since different dark energy models may predict similar behavior for \(H(z)\) and \(q(z)\), the statefinder parameters become useful tools for identifying deviations from the standard \(\Lambda CDM\) cosmology.

The statefinder parameters are defined as

\begin{equation}
	r=q+2q^{2}-\frac{(1+z)}{H}\frac{dq}{dz},
	\label{35}
\end{equation}

and

\begin{equation}
	s=\frac{r-1}{3\left(q-\frac12\right)}.
	\label{36}
\end{equation}

Here, \(r\) is known as the jerk parameter and represents the third-order derivative contribution of the cosmic scale factor, while \(s\) measures the deviation of the cosmological model from the standard \(\Lambda CDM\) scenario.

The deceleration parameter used in the above expressions is given in the equation (\ref{34}) and that of constraint values in table (\ref{tab1}) The present-day statefinder parameters at \(z=0\) are obtained approximately as $r_0 \approx 0.58$ and $	s_0 \approx 0.12$ for the Pantheon\(^+\) dataset while  $r_0 \approx 0.61$ and $s_0 \approx 0.10$. The obtained negative values of \(q_0\) confirm that the Universe is presently undergoing accelerated expansion. At the same time, the values of \(r\) and \(s\) indicate that the cosmological evolution predicted by the model slightly deviates from the standard \(\Lambda CDM\) cosmology. It is well known that the fixed point corresponding to the standard cosmological constant model is $(r,s)=(1,0)$.

\begin{figure}
	\begin{minipage}{0.5\textwidth}
		\centering
	\includegraphics[scale=0.7]{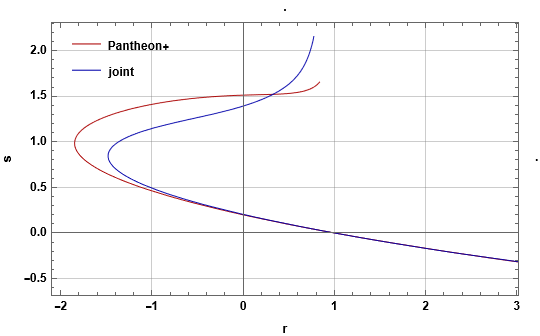}
	\caption{Behavior of the statefinder trajectory in the \((r,s)\)-plane for the present \(F(R, \Sigma ,T)\) gravity model using the observationally constrained parameter values.}\label{rs}
\end{minipage}\hfill
	\begin{minipage}{0.5\textwidth}
\centering
\includegraphics[scale=0.7]{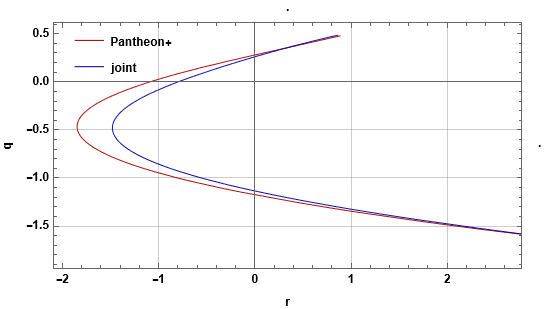}
\caption{Behavior of the statefinder trajectory in the \((r,q)\)-plane for the present \(F(R, \Sigma ,T)\) gravity model using the observationally constrained parameter values.}\label{rq}
\end{minipage}
\end{figure}

The evolutionary trajectory moves toward the neighborhood of the \((1,0)\) point, demonstrating that the model asymptotically approaches a \(\Lambda CDM\)-like behavior during the late-time evolution of the Universe (see fig. \ref{rs}). However, since the trajectories do not exactly overlap with the fixed point, the model retains a dynamical dark energy nature rather than behaving as a pure cosmological constant. For positive redshift values \((z>0)\), the statefinder pair lies away from the \(\Lambda CDM\) point, indicating stronger dynamical effects in the earlier cosmic era. As the redshift decreases and approaches the present epoch \((z=0)\), the trajectories move closer to the standard cosmological constant region. In the future evolution region \((z<0)\), the trajectories continue approaching the accelerated regime and remain stable without showing any pathological behavior. This indicates that the model predicts a smooth future expansion history.

The \((r,q)\)-plane is another important diagnostic tool that illustrates the relation between the jerk parameter and the cosmic acceleration parameter (see fig. \ref{rq}). In the early Universe, where \(q>0\), the model remains in the decelerated expansion region. During the cosmic evolution, the trajectory crosses the transition boundary at $q=0$,
which corresponds to the transition from decelerated to accelerated expansion. 
For the Pantheon\(^+\) dataset, the transition occurs approximately near $z_t \approx 0.22$, while for the joint dataset the transition takes place around $z_t \approx 0.30$. Near the present epoch, the trajectories occupy the region $q<0,\; r<1$, 
which characterizes a quintessence-like accelerating phase. As the Universe evolves toward the future region, the trajectories move further into the accelerated sector and gradually stabilize. The obtained values are compatible with several recent observational reconstructions and previously published statefinder analyses, where the present Universe generally lies near the \(\Lambda CDM\) fixed point with small deviations due to evolving dark energy contributions.

Therefore, the combined \((r,s)\) and \((r,q)\) diagnostics indicate that the proposed cosmological model successfully describes the complete cosmic evolution from the matter-dominated decelerated epoch to the present accelerated phase while remaining observationally consistent with late-time cosmological data.

\subsubsection{The equation of state parameter}

The equation of state (EoS) parameter plays a fundamental role in cosmology because it characterizes the thermodynamic nature of the cosmic fluid and determines the expansion dynamics of the Universe. In the present \(\mathcal{F}(\mathcal{R}, \Sigma ,\mathcal{T})\) gravity framework, the effective equation of state parameter is obtained as $\omega=\frac{p}{\rho}$, where the effective pressure \(p\) and energy density \(\rho\) are modified by the geometrical corrections arising from the generalized matter-curvature coupling. Using the modified field equations together with the reconstructed Hubble parametrization, the effective equation of state parameter is obtained in the following generalized form:
\begin{widetext}
\begin{equation}\label{37 }
	\omega(z)=\frac{H_0^2(-1+\chi)\Big[-D(1+z)^3(8\chi+3\lambda)\Big]}{2\Big[4(H_0-H_0\chi)^2\left(D(1+z)^3+C(1+z)^m+A(1+z)^\eta+B(1+z)^{3-3\xi}\right)(3+\lambda)\Big]}
\end{equation}
\end{widetext}
The above expression demonstrates that the evolution of the EoS parameter depends explicitly upon the Hubble expansion rate and its cosmic derivative. Consequently, the dynamical evolution of dark energy is directly influenced by the modified gravitational coupling terms present in \(\mathcal{F}(\mathcal{R}, \Sigma ,\mathcal{T})\) gravity. From observational cosmology, the equation of state parameter is one of the most important quantities used to classify different phases of the Universe. Different intervals of \(\omega\) correspond to different physical cosmic fluids, namely $\omega=1: \text{stiff matter}$, $\omega=\frac13: \text{radiation dominated era}$, $\omega=0: \text{pressureless matter era}$, $-\frac13<\omega<0: \text{quintessence-like phase}$, $\omega=-1: \Lambda CDM \ \text{cosmology}$ and $\omega<-1: \text{phantom regime}$.

Current observational investigations based on Type Ia Supernovae, Cosmic Microwave Background radiation, Baryon Acoustic Oscillations, Cosmic Chronometers, and DESI measurements strongly indicate that the present value of the EoS parameter near the cosmological constant boundary i.e. $\omega \simeq -1$, with small possible deviations toward both quintessence and phantom behavior. Also, several recent analyses have reported that observational data remain compatible with a mildly dynamical dark energy sector, where the effective EoS parameter may evolve with cosmic redshift instead of remaining exactly constant. In particular, recent studies suggest that the present value of the EoS parameter may lie approximately in the range $-1.1 \lesssim \omega_0 \lesssim -0.9$, depending on the adopted parametrization and observational dataset.  Furthermore, several modified gravity investigations and dynamical dark energy reconstructions have shown the possibility of crossing the phantom divide line $\omega=-1$, which is difficult to realize within the standard minimally coupled scalar field scenario.
\begin{figure}[ht]
	\centering
	\includegraphics[scale=0.7]{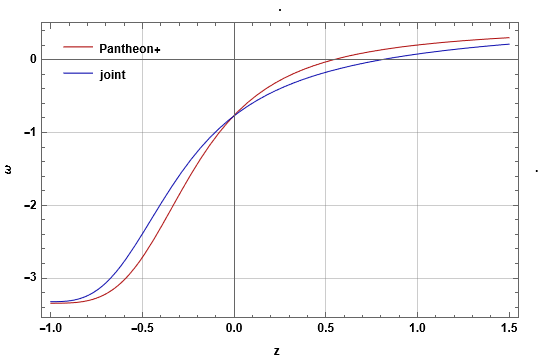}
	\caption{Evolution of the effective equation of state deceleration parameter \(q(z)\) as a function of redshift \(z\) for the constrained values of the model parameters using the Pantheon\(^+\) and joint \((CC+\mathrm{Pantheon}^+)\) datasets}.\label{w}
\end{figure}
The graphical behavior of \(\omega(z)\) indicates that the Universe evolves from a matter dominated phase toward a dark energy dominated accelerating epoch (see figure \ref{w}). At higher redshift values \((z>0)\), the equation of state parameter remains close to the pressureless matter region i.e. $\omega \approx 0$, which corresponds to the standard matter dominated cosmological era. As the Universe evolves toward lower redshift, the equation of state parameter gradually decreases and enters the negative region, indicating the onset of cosmic acceleration. Near the present epoch \((z=0)\), the model predicts values lying within the dark energy regime $-1<\omega<-\frac13$, which corresponds to a quintessence-like accelerating Universe. For suitable values of the coupling parameters \(\chi\) and \(\lambda\), the equation of state parameter may also approach the cosmological constant boundary $\omega=-1$ or even cross into the phantom region $\omega<-1$. Such phantom crossing behavior is particularly important because it cannot easily be achieved within minimally coupled scalar field cosmology. %However, generalized modified gravity theories naturally permit this transition through effective geometrical contributions.

The behavior of the effective EoS parameter in the present \(\mathcal{F}(\mathcal{R}, \Sigma ,\mathcal{T})\) gravity model indicates that the Universe evolves from a matter-dominated epoch toward a late-time accelerated dark energy dominated phase. Depending upon the values of the coupling parameters and the reconstructed Hubble function, the model may exhibit either quintessence-like behavior $(-1<\omega<-1/3)$ or phantom-like behavior $(\omega<-1)$ during the late-time cosmic evolution. The possibility of obtaining a dynamically evolving EoS parameter is one of the important advantages of generalized modified gravity theories. Unlike the standard \(\Lambda CDM\) model, where the cosmological constant remains fixed throughout the cosmic history, the present framework allows the dark energy sector to evolve naturally through geometrical corrections associated with \(\mathcal{R}\), \( \Sigma \), and \(\mathcal{T}\).

Consequently, the present \(\mathcal{F}(\mathcal{R}, \Sigma ,\mathcal{T})\) gravity model provides a physically viable description of late-time cosmic acceleration while remaining compatible with present observational constraints on the dark energy equation of state parameter.

\subsubsection{Thermodynamic interpretation}

Thermodynamics has emerged as one of the most important theoretical tools for understanding gravitational dynamics and cosmic evolution. In generalized modified gravity theories, the thermodynamic behavior of the Universe becomes especially important because the gravitational field equations themselves may be interpreted as thermodynamic relations on the cosmic horizon.

In the present \(\mathcal{F}(\mathcal{R}, \Sigma ,\mathcal{T})\) gravity framework, the FLRW Universe is treated as a thermodynamic system bounded by the apparent horizon. The apparent horizon radius is given by

\begin{equation}
	R_A=\frac{1}{H},
\end{equation}

which directly depends upon the cosmic expansion rate. The corresponding Hawking temperature associated with the apparent horizon is

\begin{equation}
	T_A=\frac{1}{2\pi R_A}
	=\frac{H}{2\pi}.
\end{equation}

This relation indicates that the temperature of the Universe evolves dynamically with the Hubble parameter. During the early Universe, where the expansion rate is extremely large, the horizon temperature remains high. However, as the Universe expands and the Hubble parameter decreases, the cosmic temperature gradually reduces.

The entropy associated with the apparent horizon is obtained through the Bekenstein-Hawking relation

\begin{equation}\label{40}
	S_A=\frac{A}{4G},
\end{equation}
where the horizon area is $A=4\pi R_A^2$. Equation (\ref{40}) reveals that the entropy is inversely proportional to the square of the Hubble parameter. 
\begin{figure}[ht]
	\centering
	\includegraphics[scale=0.7]{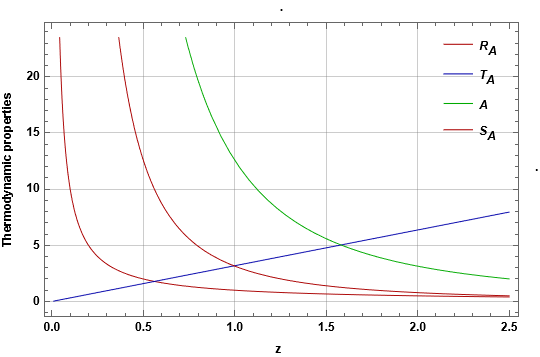}
	\caption{Thermodynamic properties parameter \(q(z)\) as a function of \(H\).}.\label{thermo}
\end{figure}

Therefore, as the Universe expands and \(H\) decreases, the entropy continuously increases. This behavior is physically consistent with the thermodynamic arrow of time and the growth of cosmic disorder during cosmic evolution. 

An important feature of the present theory is that the additional geometrical coupling terms modify the effective thermodynamic behavior of the cosmic fluid. Unlike Einstein gravity, where the entropy depends only upon the horizon area, the generalized matter-curvature couplings in \(\mathcal{F}(\mathcal{R}, \Sigma ,\mathcal{T})\) gravity contribute indirectly through the modified Hubble evolution and effective pressure-density relations. Therefore, the thermodynamic evolution of the Universe becomes strongly sensitive to the gravitational coupling parameters. This leads to richer cosmological behavior compared to standard General Relativity and allows the model to describe a dynamically evolving dark energy sector while preserving thermodynamic stability.

Overall, the thermodynamic analysis (see the figure (\ref{thermo})) confirms that the proposed \(\mathcal{F}(\mathcal{R}, \Sigma ,\mathcal{T})\) gravity model remains physically viable, thermodynamically consistent, and compatible with the observed accelerated expansion of the late-time Universe.

\section{Conclusion}

In the present investigation, we have developed a cosmological model in the framework of generalized \( \mathcal{F}(\mathcal{R},\Sigma,\mathcal{T}) \) gravity by combining nonlinear matter--geometry coupling effects with gravitationally induced particle creation and the variable generalized Chaplygin gas formalism. The analysis has been performed in the context of a homogeneous and isotropic flat FLRW Universe, where the modified gravitational interaction generates significant corrections to the conventional Einstein cosmology.

Starting from the generalized action depending upon the Ricci scalar \( \mathcal{R} \), the matter-coupling scalar \( \Sigma \), and the trace of the energy--momentum tensor \( \mathcal{T} \), the corresponding modified field equations were derived. By adopting a linear form of the gravitational functional, explicit expressions for the effective energy density and pressure were obtained in terms of the Hubble parameter and its derivative. The inclusion of particle creation modifies the standard conservation law through an effective creation pressure, thereby allowing the Universe to behave as an open thermodynamic system. This mechanism contributes naturally to the accelerated expansion without introducing additional exotic scalar fields.

The cosmological dynamics were reconstructed through an exact analytical expression of the Hubble parameter obtained from the combined contributions of particle creation matter and variable generalized Chaplygin gas fluid. The resulting Hubble parametrization was subsequently constrained using the Pantheon\(^+\) supernova compilation and the combined observational Hubble plus Pantheon\(^+\) datasets. The statistical analysis demonstrated that the obtained best-fit values remain fully compatible with current late-time cosmological observations. In particular, the estimated values of the Hubble constant are consistent with recent observational determinations based on low-redshift probes.

The physical behavior of the model was examined through several cosmological diagnostics. The evolution of the deceleration parameter confirmed the existence of a smooth transition from an early decelerated matter-dominated epoch toward the presently accelerating phase of the Universe. The obtained transition redshift lies within the observationally accepted interval reported in recent cosmological reconstructions. Moreover, the future behavior of the deceleration parameter indicates the persistence of accelerated expansion during the later stages of cosmic evolution.

To further distinguish the present framework from other dark energy scenarios, the statefinder diagnostic pair \((r,s)\) together with the \((r,q)\)-plane analysis were investigated. The trajectories in the statefinder plane approach the neighborhood of the standard \(\Lambda CDM\) fixed point \((r,s)=(1,0)\), indicating that the proposed model successfully reproduces a cosmological constant dominated behavior at late times while still preserving the dynamical nature of dark energy. The \((r,q)\)-plane also demonstrates the transition from the decelerated region to the accelerated expansion regime in a physically stable manner.

The effective equation of state parameter exhibits a dynamically evolving behavior throughout cosmic evolution. The obtained evolution shows that the Universe passes naturally from a pressureless matter dominated phase to a quintessence-like accelerating era. Depending upon the values of the coupling parameters, the model may also approach or cross the phantom divide line, which is a characteristic feature of several generalized modified gravity theories. Such behavior provides additional flexibility in explaining the observational indications of evolving dark energy.

The thermodynamic properties of the model were also analyzed within the apparent horizon framework. The evolution of the horizon temperature and entropy indicates that the generalized second law of thermodynamics remains satisfied during the cosmic evolution. The entropy growth associated with the apparent horizon confirms the thermodynamic consistency of the present gravitational framework. Furthermore, the additional matter--geometry coupling terms modify the thermodynamic evolution of the Universe and enrich the overall cosmological behavior compared with standard general relativity.

The combined observational and theoretical analysis therefore suggests that the present \( \mathcal{F}(\mathcal{R},\Sigma,\mathcal{T}) \) gravity model with particle creation represents a physically consistent and observationally viable description of the late-time Universe. The framework successfully explains the transition from deceleration to acceleration, supports dynamically evolving dark energy behavior, and preserves thermodynamic stability within a generalized gravitational scenario.

Finally, the present work may be extended in several directions. Future investigations involving baryon acoustic oscillation data, cosmic microwave background observations, growth rate analysis, and higher-order cosmographic diagnostics may provide stronger constraints on the coupling parameters of the theory. In addition, stability analysis, perturbation behavior, and reconstruction techniques in anisotropic cosmological backgrounds may further clarify the physical implications of the proposed gravitational framework.

\section*{Authors contributions}
NM: Writing-review \& editing, Validation, Resources, Methodology. SHS: Writing-original draft, Software, Methodology, Investigation, Conceptualization. AP: Validation, Software, Investigation, Conceptualization. MZ: Validation, Software, Visualization, Supervision, Funding acquisition. 

\section*{Data availability}
No datasets were generated or analyzed during the current study.

\section*{Competing interests} The authors declare no competing interests.

%\section*{Ethics Declaration} This article does not contain any studies involving human participants or animals performed by any of the authors.

\section*{Acknowledgments}
This research is funded by the Science Committee of the Ministry of Science and Higher Education of the Republic of Kazakhstan (Grant No. AP23483654). The IUCAA, Pune, India, is acknowledged by the author S. H. Shekh \& A. Pradhan for giving the facility through the Visiting Associateship programmes.

%\newpage 
\appendix
\begin{center}
	\textbf{Appendix of 3.2}
\end{center}
Derivation the corresponding expression for the Hubble parameter \(H(z)\).
As we have $$\rho_{m}=\rho_{m0}(1+z)^{3(1-\zeta)}$$
and $$\rho_{vg}=\rho_{vg0}\left[A_{s}(1+z)^{m}+(1-A_{s})(1+z)^{3(1+\alpha)}\right]^{\frac{1}{1+\alpha}}$$
The total energy density $\rho=\rho_{vg}+\rho_{m}$, becomes
\begin{widetext}
\begin{align}
	\rho=
	\rho_{m0}(1+z)^{3(1-\zeta)}
	+\rho_{vg0}
	\left[
	A_{s}(1+z)^{m}
	+(1-A_{s})(1+z)^{3(1+\alpha)}
	\right]^{\frac{1}{1+\alpha}}.
	\label{total_density}
\end{align}
\end{widetext}
Substituting Eq.~\eqref{total_density} into Eq.~\eqref{11}, we obtain
\begin{widetext}
\begin{align}
	\rho_{m0}(1+z)^{3(1-\zeta)}
	+\rho_{vg0}
	\left[
	A_{s}(1+z)^{m}
	+(1-A_{s})(1+z)^{3(1+\alpha)}
	\right]^{\frac{1}{1+\alpha}}
	\nonumber\\
	=
	\frac{
		3\lambda_{4}(1-\chi)^{2}H^{2}
		+3\lambda_{2}(\chi-1)(\dot{H}+H^{2})
	}{
		\lambda_{1}\lambda_{4}+\lambda_{2}\lambda_{3}
	}.
	\label{substitute_density}
\end{align}
\end{widetext}
Multiplying both sides by \((\lambda_{1}\lambda_{4}+\lambda_{2}\lambda_{3})\), we get
\begin{widetext}
\begin{align}
	(\lambda_{1}\lambda_{4}+\lambda_{2}\lambda_{3})
	\Bigg[
	\rho_{m0}(1+z)^{3(1-\zeta)}
	+\rho_{vg0}
	\left(
	A_{s}(1+z)^{m}
	+(1-A_{s})(1+z)^{3(1+\alpha)}
	\right)^{\frac{1}{1+\alpha}}
	\Bigg]
	\nonumber\\
	=
	3\lambda_{4}(1-\chi)^{2}H^{2}
	+3\lambda_{2}(\chi-1)(\dot{H}+H^{2}).
	\label{expanded_eq}
\end{align}
\end{widetext}
Using the relation between cosmic time and redshift,
\begin{align}
	\dot{H}=\frac{dH}{dt}
	=
	\frac{dH}{dz}\frac{dz}{dt},
\end{align}
and since
\begin{align}
	\frac{dz}{dt}=-(1+z)H,
\end{align}
we obtain
\begin{align}
	\dot{H}=-(1+z)H\frac{dH}{dz}.
	\label{hdot_relation}
\end{align}
Substituting Eq.~\eqref{hdot_relation} into Eq.~\eqref{expanded_eq}, we obtain
\begin{widetext}
\begin{align}
	(\lambda_{1}\lambda_{4}+\lambda_{2}\lambda_{3})
	\Bigg[
	\rho_{m0}(1+z)^{3(1-\zeta)}
	+\rho_{vg0}
	\left(
	A_{s}(1+z)^{m}
	+(1-A_{s})(1+z)^{3(1+\alpha)}
	\right)^{\frac{1}{1+\alpha}}
	\Bigg]
	\nonumber\\
	=
	3\lambda_{4}(1-\chi)^{2}H^{2}
	+3\lambda_{2}(\chi-1)
	\left[
	-(1+z)H\frac{dH}{dz}+H^{2}
	\right].
	\label{hz_equation}
\end{align}
\end{widetext}
Expanding the right-hand side, we get
%\begin{widetext}
%\begin{align}(\lambda_{1}\lambda_{4}+\lambda_{2}\lambda_{3})\Bigg[\rho_{m0}(1+z)^{3(1-\zeta)}+\rho_{vg0}\left(A_{s}(1+z)^{m}+(1-A_{s})(1+z)^{3(1+\alpha)}\right)^{\frac{1}{1+\alpha}}\Bigg]	\nonumber\\	=3\Big[\lambda_{4}(1-\chi)^{2}+\lambda_{2}(\chi-1)\Big]H^{2}-3\lambda_{2}(\chi-1)(1+z)H\frac{dH}{dz}.	\label{differential_eq_H}\end{align}\end{widetext}
Thus, the differential equation governing the Hubble parameter becomes
\begin{widetext}
\begin{align}
	(1+z)H\frac{dH}{dz}
	-
	\frac{
		\lambda_{4}(1-\chi)^{2}
		+\lambda_{2}(\chi-1)
	}{
		\lambda_{2}(\chi-1)
	}H^{2}
	\nonumber\\
	=
	-\frac{
		(\lambda_{1}\lambda_{4}+\lambda_{2}\lambda_{3})
	}{
		3\lambda_{2}(\chi-1)
	}
	\Bigg[
	\rho_{m0}(1+z)^{3(1-\zeta)}
	+\rho_{vg0}
	\left(
	A_{s}(1+z)^{m}
	+(1-A_{s})(1+z)^{3(1+\alpha)}
	\right)^{\frac{1}{1+\alpha}}
	\Bigg].
	\label{final_diff_eq}
\end{align}
\end{widetext}
Starting from the first-order differential equation obtained for the Hubble parameter,
\begin{equation}
	\frac{dY}{dz}+P(z)Y=Q(z),
	\label{eq:linear}
\end{equation}
where $Y(z)=H^2(z)$, and $P(z)=-\frac{\eta}{1+z}$,
the equation (\ref{eq:linear}) becomes 
\begin{equation}
	\frac{dY}{dz}-\frac{\eta}{1+z}Y=Q(z).
	\label{eq:mainode}
\end{equation}
The integrating factor corresponding to Eq.~(\ref{eq:mainode}) is
\begin{equation}
	\mu(z)=\exp\left[\int -\frac{\eta}{1+z}dz\right]
	=(1+z)^{-\eta}.
\end{equation}
Multiplying Eq.~(\ref{eq:mainode}) by the integrating factor gives
\begin{equation}
	(1+z)^{-\eta}\frac{dY}{dz}
	-\frac{\eta}{1+z}(1+z)^{-\eta}Y
	=(1+z)^{-\eta}Q(z).
\end{equation}
The left-hand side reduces to
\begin{equation}
	\frac{d}{dz}\left[(1+z)^{-\eta}Y\right]
	=(1+z)^{-\eta}Q(z).
\end{equation}
Integrating both sides,
\begin{equation}
	(1+z)^{-\eta}Y
	=
	\int (1+z)^{-\eta}Q(z)\,dz+C,
\end{equation}
which yields
\begin{equation}
	H^2(z)
	=
	(1+z)^{\eta}
	\left[
	C+\int (1+z)^{-\eta}Q(z)\,dz
	\right].
	\label{eq:hgeneral}
\end{equation}
Substituting the explicit form of \(Q(z)\),
\begin{align}
	Q(z)=
	-\frac{2(\lambda_1\lambda_4+\lambda_2\lambda_3)}
	{3\lambda_2(\chi-1)(1+z)}
	\Bigg[
	&\rho_{m0}(1+z)^{3(1-\zeta)}
	\nonumber\\
	&+
	\rho_{vg0}
	\left(
	A_s(1+z)^m
	+
	(1-A_s)(1+z)^3
	\right)
	\Bigg],
\end{align}
we obtain
\begin{widetext}
\begin{align}
	H^2(z)
	=
	(1+z)^{\eta}
	\Bigg[
	C
	-
	\frac{2(\lambda_1\lambda_4+\lambda_2\lambda_3)}
	{3\lambda_2(\chi-1)}
	\int
	(1+z)^{-(\eta+1)}
	\Bigg(
	&\rho_{m0}(1+z)^{3(1-\zeta)}
	\nonumber\\
	&+
	\rho_{vg0}
	\left[
	A_s(1+z)^m
	+
	(1-A_s)(1+z)^3
	\right]
	\Bigg)
	dz
	\Bigg].
\end{align}
\end{widetext}
Separating the integral,
\begin{equation}
	H^2(z)
	=
	(1+z)^{\eta}
	\left[
	C+I_1+I_2
	\right],
\end{equation}
where
\begin{equation}
	I_1
	=
	-\frac{2(\lambda_1\lambda_4+\lambda_2\lambda_3)\rho_{m0}}
	{3\lambda_2(\chi-1)}
	\int
	(1+z)^{3(1-\zeta)-\eta-1}dz,
\end{equation}
and
\begin{align}
	I_2
	=
	-\frac{2(\lambda_1\lambda_4+\lambda_2\lambda_3)\rho_{vg0}}
	{3\lambda_2(\chi-1)}
	\int
	(1+z)^{-(\eta+1)}
	\Big[
	&A_s(1+z)^m
	\nonumber\\
	&+
	(1-A_s)(1+z)^3
	\Big]dz.
\end{align}
Evaluating \(I_1\),
\begin{equation}
	I_1
	=
	-\frac{
		2(\lambda_1\lambda_4+\lambda_2\lambda_3)\rho_{m0}
	}
	{
		3\lambda_2(\chi-1)\,[3(1-\zeta)-\eta]
	}
	(1+z)^{3(1-\zeta)-\eta}.
\end{equation}
Now, considering the simplified case for $\alpha$, %\(\alpha=0\), 
the viscous generalized Chaplygin gas density becomes
\begin{equation}
	\rho_{vg}
	=
	\rho_{vg0}
	\left[
	A_s(1+z)^m
	+
	(1-A_s)(1+z)^3
	\right].
\end{equation}
Hence,
\begin{align}
	I_2
	=
	-\frac{
		2(\lambda_1\lambda_4+\lambda_2\lambda_3)\rho_{vg0}
	}
	{
		3\lambda_2(\chi-1)
	}
	\Bigg[
	&A_s
	\int
	(1+z)^{m-\eta-1}dz
	\nonumber\\
	&+
	(1-A_s)
	\int
	(1+z)^{2-\eta}dz
	\Bigg].
\end{align}
Evaluating the integrals,
\begin{equation}
	\int
	(1+z)^{m-\eta-1}dz
	=
	\frac{(1+z)^{m-\eta}}{m-\eta},
\end{equation}
and
\begin{equation}
	\int
	(1+z)^{2-\eta}dz
	=
	\frac{(1+z)^{3-\eta}}{3-\eta}.
\end{equation}
Thus,
\begin{align}
	I_2
	=
	-\frac{
		2(\lambda_1\lambda_4+\lambda_2\lambda_3)\rho_{vg0}
	}
	{
		3\lambda_2(\chi-1)
	}
	\left[
	\frac{A_s(1+z)^{m-\eta}}{m-\eta}
	+
	\frac{(1-A_s)(1+z)^{3-\eta}}{3-\eta}
	\right].
\end{align}
Substituting \(I_1\) and \(I_2\) into the solution,
\begin{widetext}
\begin{align}
	H^2(z)
	=
	&C(1+z)^\eta
	\nonumber\\
	&-
	\frac{
		2(\lambda_1\lambda_4+\lambda_2\lambda_3)\rho_{m0}
	}
	{
		3\lambda_2(\chi-1)[3(1-\zeta)-\eta]
	}
	(1+z)^{3(1-\zeta)}
	\nonumber\\
	&-
	\frac{
		2(\lambda_1\lambda_4+\lambda_2\lambda_3)\rho_{vg0}
	}
	{
		3\lambda_2(\chi-1)
	}
	\left[
	\frac{A_s(1+z)^m}{m-\eta}
	+
	\frac{(1-A_s)(1+z)^3}{3-\eta}
	\right].
	\label{eq:h2before}
\end{align}
\end{widetext}
Applying the boundary condition
\begin{equation}
	H(z=0)=H_0,
\end{equation}
we obtain
\begin{align}
	C
	=
	&H_0^2
	\nonumber\\
	&+
	\frac{
		2(\lambda_1\lambda_4+\lambda_2\lambda_3)\rho_{m0}
	}
	{
		3\lambda_2(\chi-1)[3(1-\zeta)-\eta]
	}
	\nonumber\\
	&+
	\frac{
		2(\lambda_1\lambda_4+\lambda_2\lambda_3)\rho_{vg0}
	}
	{
		3\lambda_2(\chi-1)
	}
	\left[
	\frac{A_s}{m-\eta}
	+
	\frac{1-A_s}{3-\eta}
	\right].
\end{align}
Substituting this constant back into Eq.~(\ref{eq:h2before}),
\begin{widetext}
\begin{align}
	H^2(z)
	=
	&H_0^2(1+z)^\eta
	\nonumber\\
	&+
	\frac{
		2(\lambda_1\lambda_4+\lambda_2\lambda_3)\rho_{m0}
	}
	{
		3\lambda_2(\chi-1)[3(1-\zeta)-\eta]
	}
	\left[
	(1+z)^\eta-(1+z)^{3(1-\zeta)}
	\right]
	\nonumber\\
	&+
	\frac{
		2(\lambda_1\lambda_4+\lambda_2\lambda_3)\rho_{vg0}
	}
	{
		3\lambda_2(\chi-1)
	}
	\Bigg[
	\left(
	\frac{A_s}{m-\eta}
	+
	\frac{1-A_s}{3-\eta}
	\right)
	(1+z)^\eta
	\nonumber\\
	&\hspace{4cm}
	-
	\left(
	\frac{A_s(1+z)^m}{m-\eta}
	+
	\frac{(1-A_s)(1+z)^3}{3-\eta}
	\right)
	\Bigg].
\end{align}
\end{widetext}
Now defining the dimensionless constants

\begin{equation}
	B=\Omega_{m0},
\end{equation}
\begin{equation}
	C=\Omega_{vg0}\left(\frac{A_s}{m-\eta}\right),
\end{equation}
and
\begin{equation}
	D=\Omega_{vg0}\left(\frac{1-A_s}{3-\eta}\right),
\end{equation}
the dimensionless Hubble parameter becomes
\begin{widetext}
\begin{equation}
	H^2(z)
	=H_0 \sqrt{A(1+z)^\eta-B(1+z)^{3(1-\zeta)}-C(1+z)^m-D(1+z)^3}.
\end{equation}
\end{widetext}

%Differentiating the entropy relation with respect to cosmic time gives
%\begin{equation}\dot S_A=-\frac{2\pi\dot H}{GH^3}.\end{equation}
% For an accelerating Universe satisfying \begin{equation}\dot H<0,\end{equation} the entropy production remains positive, \begin{equation}\dot S_A>0.\end{equation}
%Hence, the horizon entropy increases continuously during the late-time accelerated expansion of the Universe. The thermodynamic volume enclosed within the apparent horizon is %\begin{equation}V=\frac{4\pi}{3H^3},\end{equation} and the corresponding internal energy becomes \begin{equation}E=\rho V=\frac{4\pi\rho}{3H^3}.\end{equation} The Gibbs thermodynamic relation is expressed as \begin{equation}T_A dS_I=d(\rho V)+pdV,\end{equation} where \(S_I\) denotes the entropy of the cosmic fluid enclosed inside the apparent horizon. Using the conservation equation together with the modified Friedmann equations, the entropy variation of the internal cosmic fluid is obtained as \begin{equation}\dot S_I=\frac{1}{T_A}\left[V\dot\rho+(\rho+p)\dot V	\right].\end{equation} Consequently, the total entropy variation of the Universe is given by\begin{equation}\dot S_{tot}=\dot S_A+\dot S_I.\end{equation} The generalized second law of thermodynamics requires \begin{equation}	\dot S_{tot}\ge0.\end{equation}

\end{document}